\documentclass{emulateapj}

\shorttitle{Molecular Outflows in the Substellar Domain}
\shortauthors{Phan-Bao et al.}

\begin{document}

\title{Characterization of Molecular Outflows in The Substellar Domain}

\author{
Ngoc Phan-Bao\altaffilmark{1,2},
Chin-Fei Lee\altaffilmark{2},
Paul T.P. Ho\altaffilmark{2,3},
Cuong Dang-Duc\altaffilmark{1,4},
Di Li\altaffilmark{5}
}
\altaffiltext{1}{Department of Physics, International University-Vietnam National University HCM, 
Block 6, Linh Trung Ward, Thu Duc District, Ho Chi Minh City, Vietnam; pbngoc@hcmiu.edu.vn}
\altaffiltext{2}{Institute of Astronomy and Astrophysics, Academia Sinica, P.O. Box 23-141, Taipei 106, Taiwan, ROC; 
pbngoc@asiaa.sinica.edu.tw}
\altaffiltext{3}{Harvard-Smithsonian Center for Astrophysics, Cambridge, MA}
\altaffiltext{4}{Faculty of Physics and Engineering Physics,
University of Science-Vietnam National University HCM, 227 Nguyen Van Cu Street, District 5, Ho Chi Minh City, Vietnam}
\altaffiltext{5}{National Astronomical Observatories, Chinese Academy of Science, 
Chaoyang District Datun Rd A20, Beijing, China}

\begin{abstract}
We report here our latest search for molecular outflows from young brown dwarfs
and very low-mass stars in nearby star-forming regions. 
We have observed three sources in Taurus with the Submillimeter Array and 
the Combined Array for Research
in Millimeter-wave Astronomy at 230 GHz frequency 
to search for CO J~=~2$\rightarrow$1 outflows.
We obtain a tentative detection of a redshifted and extended gas lobe at 
about 10 arcsec from the source GM Tau, 
a young brown dwarf in Taurus with an estimated mass of 73 M$_{J}$, 
which is right below the hydrogen-burning limit.
No blueshifted emission around the brown dwarf position is detected.
The redshifted gas lobe that is elongated in the northeast direction suggests 
a possible bipolar outflow from the source with a position angle of about 36$^{\circ}$. 
Assuming that the redshifted emission is outflow emission from GM Tau, 
we then estimate a molecular outflow mass in the range from $1.9 \times 10^{-6}$~$M_{\odot}$ 
to $2.9 \times 10^{-5}$~$M_{\odot}$ 
and an outflow mass-loss rate from 
$2.7 \times 10^{-9}$~$M_{\odot}$yr$^{-1}$ to $4.1 \times 10^{-8}$~$M_{\odot}$yr$^{-1}$.
These values are comparable to
those we have observed in the young brown dwarf ISO-Oph~102 of 60 M$_{J}$ in $\rho$
Ophiuchi and the very low-mass star MHO~5 of 90~M$_{J}$ in Taurus. 
Our results suggest that the outflow process in very low-mass objects 
is episodic with duration of a few thousand years and the outflow rate 
of active episodes does not significantly change for different stages 
of the formation process of very low-mass objects. 
This may provide us with important implications
that clarify the formation process of brown dwarfs.
\end{abstract}


\section{INTRODUCTION}

Over the last nineteen years, observations of the statistical properties 
such as the initial mass function, 
velocity dispersion, multiplicity, accretion, jets 
(see \citealt{luhman07} and references therein) 
of brown dwarfs (BD) (13--75~$M_{\rm J}$)
and very low-mass (VLM) stars (0.1--0.2 $M_{\odot}$)
(hereafter, VLM objects) in nearby star-forming
regions have shown that all these properties of VLM objects
form a continuum with those of low-mass stars.
These observations therefore strongly support the starlike models 
(see \citealt{whitworth} and references therein)
that VLM objects form in the same manner as low-mass stars.
While the starlike models predict that pre-BD cores, which are 
produced by turbulent fragmentation
of molecular clouds \citep{padoan} or gravitational fragmentation
\citep*{bonnell},
are dense enough to be gravitationally unstable,
it is still unclear how the physical processes of BD formation occur at later stages,
such as at class 0, I and II. 

Since bipolar molecular outflows, which are ambient gas swept up 
by an underlying jet/wind (see \citealt{bachiller,mckee} and references therein), 
are a basic component of the star formation process,
studying the molecular outflow properties will therefore help us
understand the BD formation mechanism.
In the last few years, we have reported the first detections of
bipolar molecular outflows from the class II BD ISO-Oph~102 
in $\rho$ Ophiuchi \citep{pb08}, the class II VLM star MHO~5 in Taurus \citep{pb11}.
Our estimated values of outflow mass and mass loss rate in these VLM objects 
are over an order of magnitude smaller than the typical values in low-mass stars. 
These results have implied that the outflow process in VLM objects is a
scaled-down version of that in low-mass stars.
Our detections have also provided strong observational constraints such as 
velocity, size, mass and mass-loss rate of the outflow process for 
the simulation of BD formation (e.g., \citealt*{mach}). 
Although the molecular outflow process holds important clues to BD
formation, however, only four detections of molecular outflows in the substellar
domain have been reported so far:
two molecular outflows from class 0/I proto BD and VLM candidates
(L1014-IRS: \citealt{bourke}; L1148-IRS: \citealt{kau}),
one from a class II BD (ISO-Oph 102, $\sim$60~$M_{\rm J}$: \citealt{pb08}) 
and one from a class II VLM star (MHO 5, $\sim$90~$M_{\rm J}$: \citealt{pb11}).

In this paper, we present our millimeter observations of
three class II BDs in Taurus. 
Sec.~2 presents our sample,
Sec.~3 reports our millimeter observations and the data reduction,
Sec.~4 presents the observational results, Sec.~5 discusses the outflow process and the formation
mechanism of VLM objects, Sec.~6 summarizes our results.
\section{SAMPLE SELECTION}
Our sample consists of eight targets 
in $\rho$ Ophiuchi and Taurus. All of them are class II VLM objects.
The observations of five of them have been reported in \citet{pb08,pb11}. 

In this paper, we present the observations of 
three remaining targets
in Taurus (147~parsecs, see \citealt{loi}): 
GM~Tau, 2MASS~J041411.88+2811535 (hereafter 2M 0414) and 
2MASS~J043814.86+2611399 (hereafter 2M 0438). 
GM Tau was identified as a pre-main-sequence star \citet{briceno93}, in 
the dark cloud HCL~2 or TMC~1 (e.g., \citealt{gold}) according to its position. 
The source was then spectroscopically classified as an M6.5 dwarf 
with an estimated mass of $\sim$73~M$_{\rm J}$ \citep{white03}.
GM Tau shows an obvious P Cygni profile (see Figure~4 in \citealt{white03})
with blue-shifted absorption components superposed on the H$\alpha$ accretion
emission profile, which strongly indicates a mass-loss process as seen in higher mass
T Tauri stars. 
2M~0414 is an M6.25 dwarf of 75~M$_{\rm J}$ \citep{lu04,muz05},
the source lies in the dark cloud LDN~1495 
(e.g., \citealt{gold}) according to its position. 
The H$\alpha$ emission of 2M~0414 also shows a clear P Cygni profile 
(see Figure~4 in \citealt{muz05}), implying a mass-loss process occurring in the
source.
2M 0438 is an M7.25 dwarf of 70~M$_{\rm J}$ \citep{lu04,muz05}, 
which lies in the same dark cloud as GM~Tau.
This BD exhibits strong forbidden emission lines (FELs) \citep{lu04} as seen in MHO~5
that could be associated with outflow activities. 
They are therefore excellent
targets for our search for molecular outflows in VLM objects.
One should note that no detection of any cores associated with the three sources
has been reported so far.
\section{OBSERVATIONS AND DATA REDUCTION}
We observed GM~Tau with the Submillimeter Array (SMA), and 2M~0414 and 2M~0438 with Combined Array
for Research in Millimeter-wave Astronomy (CARMA). The
observing log of the three young BDs is given in Table~\ref{log}.
\subsection{SMA observations}
The SMA\footnote{
The Submillimeter Array is a joint project between the 
Smithsonian Astrophysical Observatory and 
the Academia Sinica Institute of Astronomy and Astrophysics 
and is funded by the Smithsonian Institution and the Academia Sinica.} 
receiver band at 230~GHz (see \citealt{ho}) 
was used for the observations of GM~Tau on 2010 October 26. 
Zenith opacities at 225 GHz were typically in the range 0.1--0.16.
Both 4 GHz-wide sidebands, which are separated
by 8 GHz, were used. 
The SMA correlator was configured with a high spectral
resolution of 0.2~MHz ($\sim$0.27~km~s$^{-1}$) per channel for $^{12}$CO, $^{13}$CO, 
and C$^{18}$O $J=2 \rightarrow 1$ lines.
For the remainder of each sideband, we set up a lower resolution of 3.25~MHz per channel. 
We used quasars 3C~111 and 3C~273 for gain and passband calibration
of GM~Tau, respectively. Uranus was observed for flux calibration for the target. 
The uncertainty in the absolute flux calibration is $\sim$10\%. 

We reduced and further analyzed the data with the MIR software package
and the MIRIAD package adapted for the SMA, respectively. 
All eight SMA antennas were operated in the compact 
configuration, resulting in a synthesized beam of 
3$''$.05~$\times$~2$''$.82 
with a position angle of 61$^{\circ}$ (natural weighting). 
The FWHM of the primary beam is about 50$''$~at the observed frequencies. 
The rms sensitivity was $\sim$1~mJy for the continuum 
and $\sim$0.15~Jy~beam$^{-1}$ per channel for the line data (Table~\ref{log}).
\subsection{CARMA observations}
We observed the two BDs 2M~0414 and 2M~0438 
with CARMA at 230 GHz in 2010 August. 
All six 10.4 m, and nine 6.1 m antennas were
operated in the D configuration.
Zenith opacities at 227 GHz were in
the range 0.21$-$0.3 and 0.23$-$0.27 for 
2M 0414 and 2M 0438, respectively.
All eight 500 MHz-wide bands (a maximum bandwidth of 4 GHz per sideband), 
which may be positioned independently with the IF bandwidth, 
were used with different spectral resolutions for $^{12}$CO~$J=2 \rightarrow 1$ line.
The eight bands were set up in the following modes with the 2-BIT level: 
8 MHz, 31 MHz and 62 MHz with 383 channels per band; 
125 MHz, 250 MHz and 500 MHz with 319, 191 and 95 channels per band, respectively.
These modes give a wide range of spectral resolutions from 0.03 to 6.8 km~s$^{-1}$
at 230 GHz.
We observed quasar 3C 111 for gain calibration, 3C 84 and 3C 454.3 for passband calibration, 
and Uranus and 3C 84 for flux calibration.
The uncertainty in the absolute flux calibration is $\sim$20\%.
We reduced the data with the MIRIAD package adapted for the CARMA. 
The synthesized beam sizes are about 2$''$.09~$\times$~1$''$.69 
and 2$''$.19~$\times$~1$''$.68 using natural weighting
for 2M~0414 and 2M~0438, respectively.
The FWHM of the primary beam for the 10.4$\times$6.1 m antennas is about 36$''$ at 230 GHz.
The rms sensitivities of the continuum and the line data are listed in Table~\ref{log}.
\section{RESULTS}
\subsection{2M 0414 and 2M 0438}
2M 0414 and 2M 0438 are strong accretors 
(see Table~\ref{comp} and references therein).
While the H$\alpha$ emission with an obvious P Cygni profile
observed in 2M~0414 \citep{muz05} indicates 
an outflow process and the presence of FELs
in 2M~0438 \citep{lu04} suggests outflow activities,
our carbon monoxide (CO $J=2 \rightarrow 1$) maps from 
CARMA data however do not reveal any outflows from 2M~0414 and 2M~0438.
The non-detection of molecular outflows in these two BDs indicates three possible scenarios
as discussed in \citet{pb11}:
(1) The outflow process has already stopped;
(2) There is not much gas surrounding the sources;
(3) The outflow process in these BDs is too weak to
be detectable at the millimeter wavelengths. 

For the case of 2M~0414, the outflow process has clearly been indicated
in its H$\alpha$ emission, therefore, the first scenario should be ruled
out. As the source is located in the dense region of $^{12}$CO and 
$^{13}$CO $J=1 \rightarrow 0$ of LDN~1495 (see Figure~4 in \citealt{gold}),
the second scenario is thus unlikely in this case. 
Finally, the bolometric luminosity of 2M~0414 is 0.015~$L_{\odot}$
\citep{lu04}, about 3 times less luminous than that of GM~Tau 
(0.047~$L_{\odot}$, \citealt{lu04}).
If we assume that the outflow force vs. bolometric luminosity
correlation of proto stars (e.g., \citealt{taka}) 
is applicable for young BDs, 
this thus indicates that 
the outflow force of 2M~0414 would be weaker 
(i.e., weaker molecular outflow emission)
than that of GM~Tau. 
As a possible outflow from GM~Tau is marginally detected (see Section~4.2),
therefore, the third scenario would be a reasonable explanation for 
the non-detection of outflows from 2M~0414.

For the case of 2M~0438, 
the source is also located in the dense region (see Figure~4 in \citealt{gold})
and it is close to GM~Tau (at a distance of only $\sim$2.8$'$), 
the second scenario is thus very unlikely.  
The FELs may be associated with outflow or accretion activities,
the first scenario is thus still possible.
One should note that the bolometric luminosity of 2M~0438
is very low (0.0018~$L_{\odot}$, \citealt{lu04}),
therefore, the outflow force is expected to be much less powerful 
than that of any class II BDs with molecular outflows detected so far.
So, the first and the third scenarios are both possible for 2M~0438. 
 
One should also note that we did not detect the dust continuum emission from 2M~0414 and 
2M~0438 with an upper limit of about $1\sigma$ (1$\sigma$ = 1.4 mJy for 2M~0438 and 1.1~mJy for 2M~0414). 
Our measurements are comparable within error bars to those reported in \citet{scholz06},
0.91$\pm$0.65~mJy for 2M~0414 and 2.29$\pm$0.75~mJy for 2M~0438. 

\subsection{GM Tau}
GM Tau is also a strong accretor with a mass accretion rate stronger than that
of 2M~0414 and 2M~0438 (Table~\ref{comp}). 
Its obvious P Cygni profile of H$\alpha$ emission strongly indicates
the outflow process occurring in the BD.

\begin{figure}
\vskip 1in
\hskip -0.25in
\centerline{\includegraphics[width=4in,angle=0]{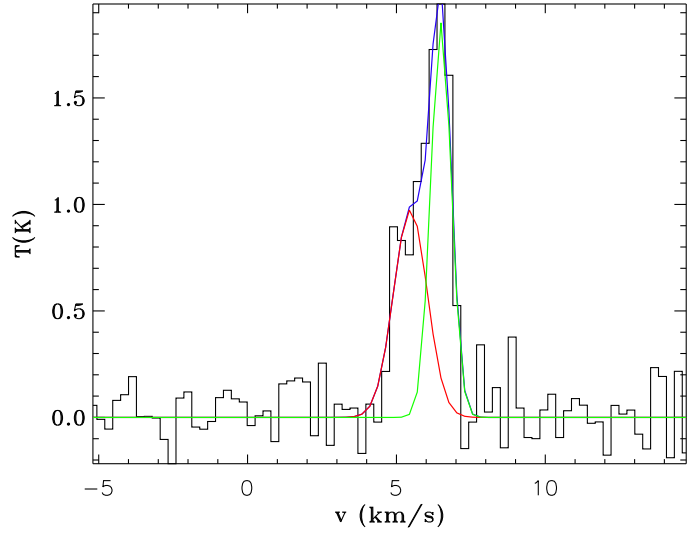}}
\caption{\normalsize 
The $^{13}$CO $J=1 \rightarrow 0$ spectrum toward GM Tau overlaid with two Gaussian components,
5.5~km~s$^{-1}$ (red line) and 6.5~km~s$^{-1}$ (green line).
The data were taken from \citet{qian12}, which was part of the FCRAO survey 
by \citet{gold} and reprocessed with a better
deconvolution. The FWHM of the FCRAO primary beam is about 50\arcsec.
\label{vsys}}
\end{figure}

\begin{figure}
\vskip 1in
\hskip -0.25in
\centerline{\includegraphics[width=5in,angle=-90]{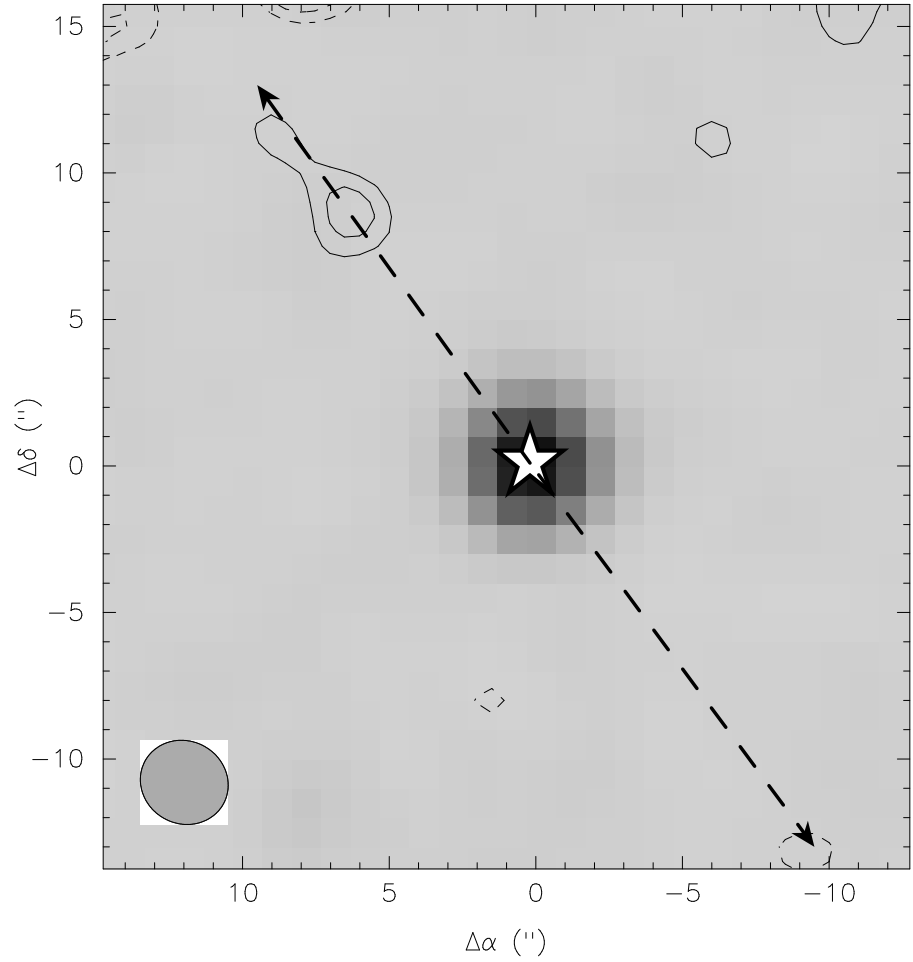}}
\caption{\normalsize Overlay of the J-band (1.25 $\mu$m) near-infrared 
Two Micron All Sky Survey image and
the integrated intensity in the CO $J=2 \rightarrow 1$ line 
emission from 6.6 to 7.2~km~s$^{-1}$ line-of-sight velocities towards GM~Tau.
The contours are $-$4, $-$3, 3, 4,...times the 
rms of 0.05~Jy~beam$^{-1}$~km~s$^{-1}$. The source is visible in the near-infrared image. 
The position angle of the outflow is about 36$^{\circ}$. 
The expected outflow direction is indicated by the arrows.
The star symbol represents the BD position. 
The synthesized beam is shown in the bottom
left corner.
\label{map}}
\end{figure}

The systemic velocity of GM Tau has not been available in the literature so far. 
To measure the systemic velocity of the BD,  
we extract a $^{13}$CO $J=1 \rightarrow 0$ spectrum toward GM~Tau with 
the reprocessed FCRAO Taurus survey data from \citet{qian12}.
The FWHM of the $^{13}$CO line is about 2.2~km~s$^{-1}$ (Fig.~\ref{vsys}), 
which is better described by two Gaussian components with peak velocities of 
$5.5\pm0.3$~km~s$^{-1}$ and $6.5\pm0.3$~km~s$^{-1}$, respectively. 
The large-scale velocity gradient of $^{13}$CO seems to roughly follow a
Larson’s law-type power-law with respect to spatial scales (see Fig. 18 in \citealt{qian12}).  
The FWHM of the FCRAO primary beam is about 50$''$.
Within 50$''$, it is thus normal in Taurus to have gas components moving
at 0.5$-$1~km~s$^{-1}$ with respect to each other. 
The FCRAO data therefore suggests the systemic velocity of GM~Tau to be
in the range from 5.5 to 6.5~km~s$^{-1}$. 
We thus take an average value of 6.0~km~s$^{-1}$ for the systemic velocity of 
the BD.
Figure~\ref{map} presents the integrated intensity in the CO~$J=2 \rightarrow 1$ 
emission towards GM~Tau. Our map reveals only a redshifted
($\sim$6.6$-$7.2~km~s$^{-1}$) CO gas lobe around the BD position.
No blueshifted emission is detected.  
It is therefore difficult to confirm that the 
redshifted lobe is outflow emission from GM Tau.
However, the gas lobe is elongated in the northeast direction of the BD position,
with a position angle of about 36$\degr$.
This suggests that the redshifted emission is possibly outflow emission from GM~Tau.
One should note that the redshifted component
is marginally detected at only $4\sigma$.
Therefore, if it really comes from an outflow of the BD, 
the detection of the CO outflow from GM~Tau should be considered as a tentative detection.
Our previous detections \citep{pb08,pb11} have shown that
the molecular outflow in VLM objects is bipolar
as seen in low-mass stars and the intensity of the outflow emission
differs significantly between the redshifted and blueshifted components.
Therefore, the non-detection of blueshifted outflow emission
from GM~Tau is probably due to that the emission is too weak to be detected with SMA.
The redshifted emission is brighter than the blueshifted 
emission implies that the redshifted jet propagates
into denser gas than the blueshifted jet, leading to a larger swept up mass of CO gas (i.e., stronger
molecular outflow emission).
Deeper observations are needed to confirm this scenario.

If the detected redshifted gas lobe is a component of outflows from GM~Tau, 
we then follow the standard manner \citep{cabrit,andre} as used in the
previous papers \citep{pb08,pb11} to calculate the
outflow properties.
The size of the redshifted CO gas lobe is about 5$''$ corresponding
to $\sim$700~AU in length (see Figure~\ref{map}). 
We assume a value of 20~K 
for the excitation temperature, we then derive a lower limit to 
the outflow mass $M_{\rm out} \sim 1.9 \times 10^{-6} M_{\odot}$. 
The correction factors due to
optical depth and missing flux for the outflow mass of GM~Tau are uncertain.
However, the typical values of optical depths for class II objects
in Taurus are from 1 to 5 \citep{l88a}. 
As GM~Tau lies in the densest part of HCL~2 (see Figure~4 of \citealt{gold}),
therefore, it is reasonable to assume an optical depth of five for the case of GM~Tau.
If a missing flux factor of three
for SMA \citep{bourke} is applicable here, we then obtain an upper limit
to the outflow mass of $\sim2.9 \times 10^{-5} M_{\odot}$.

The maximum outflow velocity $v_{\rm max}$ can be computed from the observed maximum
outflow velocity and the outflow inclination.
The observed maximum outflow velocity is about 1.2 km~s$^{-1}$. 
Based on optical, near-infrared and infrared data, \citet{riaz12} 
performed disk modelling for GM~Tau with 
a circumstellar geometry consisting of a rotationally flattened infalling envelope, 
bipolar cavities, and a flared accretion disk in hydrostatic equilibrium. 
Their best-fitting model SED was obtained with the disk's inclination angle
between 70$^{\circ}$ and 80$^{\circ}$. 
We therefore use an average value of 
75$^{\circ}$ for the outflow inclination
(the angle between the outflow axis and the line of sight).
From the observed maximum outflow velocity of 1.2 km~s$^{-1}$
and the outflow inclination angle  $i = 75\degr$,
we derive the maximum outflow velocity $v_{\rm max} = 4.6$~km~s$^{-1}$. 
We then use this value to compute upper limit values for the kinematic and 
dynamic parameters. 
We find the momentum $P=8.7 \times 10^{-6}$~$M_{\odot}$~km~s$^{-1}$,
the energy $E=2.0 \times 10^{-5}$~$M_{\odot}$~km$^{2}$~s$^{-2}$.
From the outflow size of $\sim$700~AU and 
the observed maximum outflow velocity of 1.2 km~s$^{-1}$, 
we derive the dynamical time $t_{\rm dyn}$ 
for GM~Tau of about 700~yr (with a correction for the outflow inclination). 
With this dynamical time value, we then find 
the force $F=1.2 \times 10^{-8}$~$M_{\odot}$~km~s$^{-1}$~yr$^{-1}$,
and the mechanical luminosity $L=4.7 \times 10^{-6}$~L$_{\odot}$, where L$_{\odot}$
is the solar luminosity. 
If we apply a correction for the optical depth factor of five and 
the missing flux factor of three for SMA, these upper limit values 
will increase by a factor of fifteen. 
Lower limits to these parameters
could be estimated by using the outflow mass in each velocity channel
and the space velocity of the outflow, which is assumed
to be equal to the radial velocity of that channel, 
as if the gas was moving along
the line of sight \citep{cabrit}.
This could be done for observations with good signal-to-noise ratio.
However, as our detection levels of outflows in each velocity channel are rather
low, we therefore do not estimate these lower limits.

The mass-loss rate of molecular outflows can be computed by dividing the outflow mass
by the dynamical time of the outflow. 
However, the estimated dynamical time $t_{\rm dyn} \sim 700$~yr for 
the outflow from GM~Tau 
is over two orders of magnitude smaller than the age of
VLM objects in Taurus expected to be from a few $10^{5}$~yr to a few Myrs
\citep{muz03,white03}. Therefore, the correction factor of about ten
applying for the outflow dynamical time of young low-mass stars \citep*{parker}
may be not valid here. This value for VLM objects should be over a hundred 
(see further discussion in Sec.~5.1). Without correction applied for the outflow dynamical time
and using the lower and upper values of the outflow mass,
we directly derive lower and upper limits to the mass-loss rate of molecular outflows
$\dot{M}_{\rm mol} = M_{\rm out}/t_{\rm dyn}$ 
to be $2.7 \times 10^{-9}$$M_{\odot}$~yr$^{-1}$
and $4.1 \times 10^{-8}$$M_{\odot}$~yr$^{-1}$, respectively.

We now consider the possibility that the detected redshifted emission 
might be due to gravitationally bound motion
and not outflow emission. For gravitationally bound motion, 
an outflow size $l=700$~AU with a velocity $v=4.6$~km~s$^{-1}$ would require
an enclosed mass of $\geq$8.4~$M_{\odot}$ ($M \geq v^{2}l/2G$, see \citealt{lada}).
If we assume that the densest cores in Taurus have been detected
by \citet{onishi} (see their Table 2), 
we then derive a possible largest mass of $\sim$0.003~$M_{\odot}$
for a core of 700~AU in diameter. 
In addition, we also estimate an upper limit to the mass of a core of 700 AU 
using the gas density around the GM~Tau position, which is estimated from 
the FCRAO observations \citep{qian12}. 
The line intensity ratio between $^{12}$CO $J=1 \rightarrow 0$ 
and $^{13}$CO $J=1 \rightarrow 0$ is much smaller than the
corresponding isotopic ratio, thus it is safe to assume 
that the $^{12}$CO line is optically thick. 
We then derive the gas excitation temperature of about 10~K
based on the peak antenna temperature of $^{12}$CO. 
The $^{13}$CO opacity, and in turn, the gas column density can then be derived
by assuming a [$^{13}$CO]/[H$_{2}$] 
abundance ratio of $1.7 \times 10^{-6}$ \citep{frerking}. 
The resulting H$_{2}$ column density is $6.0 \times 10^{21}$~cm$^{-2}$, 
which is consistent with 2MASS extinction measurement on 
arcminute spatial scale \citep{qian12}. 
Using the measured H$_{2}$ column density and the FWHM of the FCRAO primary beam,
we finally obtain an upper limit of 
$\sim$0.13~$M_{\odot}$ to a core of 700~AU within 25$''$ from the GM~Tau position. 
Both the estimated masses from the \citet{onishi} observations and FCRAO are well below
the enclosed mass of 8.4~$M_{\odot}$ required
for gravitational bound motion.
We therefore conclude that the detected emission is from the outflow. 

One should note that we did not detect the dust continuum emission with an upper limit of 1~mJy
($=1\sigma$) measured at the GM~Tau position. 

\section{MOLECULAR OUTFLOWS FROM VLM OBJECTS}
\subsection{Molecular outflow properties}
So far, we have observed eight VLM objects and detected three of them having CO molecular
outflows: ISO-Oph 102 in $\rho$ Ophiuchi \citep{pb08}, 
MHO 5 \citep{pb11} and GM Tau (this paper) in Taurus. 
The molecular outflows from these three sources show
similar molecular outflow properties: small scale of 600--1000 AU, low velocity
of $<$5 km~s$^{-1}$ (with a correction of outflow inclination), 
tiny outflow mass of 10$^{-6}$--10$^{-4}$~$M_{\odot}$
and low mass-loss rate of 10$^{-9}$--10$^{-7}$~$M_{\odot}$yr$^{-1}$.
Table~\ref{mloss} lists these properties for each source.
Our molecular outflow mass estimates in VLM objects are smaller than 
the typical values 0.01$-$0.7~$M_{\odot}$ of class II low-mass stars (G, K spectral types;
see \citealt{l88b} and references therein)
by over an order of magnitude.

To estimate the mass-loss rate of the stellar wind that drives the molecular outflow in the VLM objects,
we can assume that the stellar wind-molecular gas interaction is 
momentum-conserving (e.g., \citealt{l88b,andre}).
We thus equate the momentum $P = M_{\rm out}v_{\rm max}$ of the molecular outflow 
with that supplied by the wind over the outflow's lifetime ($\dot{M}_{\rm wind}t_{\rm dyn}v_{\rm wind}$).
This yields the expression $\dot{M}_{\rm wind} = M_{\rm out}v_{\rm max}/(t_{\rm dyn}v_{\rm wind})$,
where $v_{\rm wind}$ is the wind velocity.
For ISO-Oph~102, the wind velocity is $\sim$107~km~s$^{-1}$,
which is estimated from 
$v_{\rm jet} = $~45~km~s$^{-1}$ \citep{whelan05} with a correction of outflow inclination of
$\sim65\degr$ to the line of sight \citep{pb08}.
For the cases of MHO~5 and GM~Tau, as no measurements of wind velocities have been reported so far,
so we assume a wind velocity of about 100~km~s$^{-1}$ for these two objects.
Using the upper and lower limit values of outflow mass (see Table~\ref{mloss}), 
we derive the upper and lower limits (Table~\ref{comp}) to the wind mass-loss rates
of the VLM objects, respectively.

There is a possibility that our values of wind mass-loss rates
are underestimated if the momentum from the wind
might not be transferred all to the molecular gas.
This happens when there is less molecular gas surrounding the source
at late stages (class II) than early stages (class 0, I). 
However, the wind mass-loss rates of VLM objects (e.g., ISO-Oph 102) 
estimated from molecular outflows
(see Table~\ref{mloss}) are comparable to those 
derived from optical jets using the the spectro-astrometric method 
(see Table~6 in \citealt{whelan09}, \citealt{whelan14}).
This implies that our values are not significantly underestimated.
All these values of wind mass-loss rates of VLM objects 
are smaller than a typical value 
of $\sim$10$^{-7}$$M_{\odot}$~yr$^{-1}$
for class-II low-mass stars of 0.5--5.0~$M_{\odot}$ \citep{l88a} 
by over about an order of magnitude (see Figure~\ref{comp}). 
Our results have shown that the outflow process occurs in VLM objects
is a scaled-down version of that in low-mass stars.

One should discuss here that the ratios of wind mass-loss rate to mass accretion rate 
$\dot{M}_{\rm wind}/\dot{M}_{\rm acc}$ in ISO-Oph~102, MHO~5 
(see Table~\ref{comp}) are significantly higher than that in T~Tauri stars 
($\sim$0.0003-0.4, \citealt{har}).
These high ratio values imply three possible scenarios:

(1) First, we might underestimate the accretion rates
as the accretion rates measured in the VLM objects at different epochs
may vary over an order of magnitude on timescales of months to years (e.g., \citealt*{stel}).
Therefore, the rates measured at particular epochs do not reflect
the long-term accretion rates that are more appropriate
for comparison with the outflow rates in the two VLM objects.
However, the accretion rate in MHO~5
measured at different epochs has suggested that the accretion rate
is stable, $\sim10^{-10.8}$$M_{\odot}$~yr$^{-1}$ \citep{muz03,her}.   
In the case of ISO-Oph~102, the accretion rate of $\sim10^{-9.0}$$M_{\odot}$~yr$^{-1}$ 
was measured at epoch 2003.411 \citep{natta04}, and
an increase of the accretion rate
by a factor of five within a week was also observed \citep{natta04}.
However, the accretion rate of $10^{-9.17}$$M_{\odot}$~yr$^{-1}$ \citep{gatti}
measured at epoch 2005.384 agrees with the Natta et al.'s measurement, 
suggesting a long-term accretion rate in ISO-Oph~012 of 
$\sim10^{-9.0}$$M_{\odot}$~yr$^{-1}$. 
Therefore, this scenario is unlikely the case here.

(2) Second, we might overestimate the wind mass-loss rate. 
This is due to that we did not apply a correction factor 
for the dynamical time (see Table~\ref{comp}). If the estimated
dynamical time is a lower limit to the real dynamical time, 
we then need to  apply a correction factor of about 100 for the case of ISO-Oph~102
and 1000 for MHO~5 as well as GM~Tau (see Section~5.2 for further discussion), 
instead of 10 as estimated for low-mass stars \citep{parker}, for 
the outflow dynamical time of VLM objects.
The wind mass-loss rates, hence, 
the ratios $\dot{M}_{\rm wind}/\dot{M}_{\rm acc}$
will decrease by the same factors and they are thus comparable to those
in T~Tauri stars.

(3) Third, this ratio in VLM objects is really higher than that in low-mass stars. 
This is possibly due to a sudden drop of the mass accretion rate during the formation
process of brown dwarfs as proposed by \citet{mach}. 
Further observations and theoretical works are needed to confirm these possible scenarios.

\subsection{Episodic outflows?}
The outflow dynamical times estimated for the three class-II VLM objects,
ISO-Oph 102 \citep{pb08}, MHO~5 \citep{pb11} and GM~Tau (this paper) 
are from $7.0\times10^{2}$~yr to $2.7\times10^{3}$~yr (see Table~\ref{mloss}).
These values are still about two or three orders of magnitude 
smaller than the ages of ISO-Oph~102, GM~Tau and MHO~5
expected to be from a few $10^{5}$~yr (ISO-Oph 102, \citealt{natta04})
to a few Myr (GM~Tau and MHO~5, see \citealt{muz03} and references therein).

The extreme discrepancy between the outflow dynamical time
and the age of the VLM objects  
indicates two possibilities:

(1) The extent of outflows is not completely revealed because of
the coverage and the sensitivity of our observations. 
For example, an outflow with velocity of 5~km~s$^{-1}$ and a dynamical time
of $10^{5}$~yr will have a length of about 0.5 pc, or 12$'$ at 
the distance of Taurus. 
This is more than an order of magnitude larger than the SMA primary beam size.
Therefore, any molecular outflow emission 
that lies more than half a primary beam from the source would be impossible to detect.
In addition, the full extent of outflows to the edge of the primary beam 
has not been detected (e.g., ISO-Oph 102, see Fig. 1 in \citealt{pb08}),
which is possibly due to the sensitivity limit of SMA. 
The estimated outflow dynamical time is thus a lower limit to the true outflow duration
as discussed in \citet{parker} for the case of low-mass stars.

(2) The outflow process in the VLM objects is episodic, 
occurring in class II with duration of a few $10^{3}$~yr.
An episodic outflow will have a discrete blob morphology. 
In this case, other blobs of the molecular outflows from
our VLM objects that lie outside the primary beam would not be detected.
The outflows detected by SMA are thus from the last active episodes.
If the episodicity of these outflows is confirmed, the outflow process in VLM objects will 
include quiescent and active episodes. 
One then can expect that the accretion associated with
outflow is also episodic. This is consistent with evolutionary models
as proposed by \citet{baraffe} that the accretion process in BDs at early stages 
could be episodic, including long quiescent phases of accretion
interrupted by short episodes (a few $10^{3}$ to $10^{4}$~yr) of high accretion.
Their evolutionary models taking into account episodic phases of accretion successfully
explain the significant luminosity spread observed in H-R diagrams of star-forming regions. 
 
More extended and deeper observations are therefore needed to explore 
the morphology of the outflows from these VLM objects in order to confirm the scenarios.
\subsection{Molecular outflows and the formation of VLM objects}
The outflow masses and 
the mass-loss rates of the molecular outflows from 
ISO-Oph 102, GM~Tau and MHO~5, which are class II objects,
are comparable to those of L1014-IRS \citep{bourke}, 
a proto BD candidate at a younger age (class 0/I) that has
an outflow with a mass and mass-loss rate of 
$\sim$10$^{-5}$~$M_{\odot}$
and $\sim10^{-9}$~$M_{\odot}$yr$^{-1}$, respectively.
In the case of the class I proto BD candidate L1148-IRS \citep{kau},
the outflow mass ($\sim$10$^{-3}$~$M_{\odot}$) 
and the mass-loss rate ($\sim$10$^{-7}$~$M_{\odot}$yr$^{-1}$) 
are slightly larger than that from our VLM objects 
but still comparable to the range of our lower and upper limits (see Table~\ref{mloss}).
This similarity implies that the mass-loss rate, hence the associated 
wind mass-loss and accretion rates 
in VLM objects do not significantly change in different formation stages
(from class 0/I to class II). 

As discussed in Section~5.1, 
the ratio of wind mass-loss rate to accretion rate $\dot{M}_{\rm wind}/\dot{M}_{\rm acc}$
in VLM objects is possibly higher than that in low-mass stars (Table~\ref{comp}).
If the wind mass-loss rate does not significantly change for different classes of
VLM object formation, 10$^{-11}$--10$^{-8}$~$M_{\odot}$yr$^{-1}$ (see Table~\ref{comp}), 
then, one may expect that this ratio also holds in earlier classes, such as classes 0, I.   
This suggests that the accretion rate in VLM objects expected to be also
in the range 10$^{-11}$--10$^{-9}$~$M_{\odot}$yr$^{-1}$ at earlier stages. 
The starlike models for BD formation predict that 
VLM cores produced by fragmentation \citep{padoan,bonnell} are
dense enough to be gravitationally unstable.
The detection of such a VLM core \citep*{andre12} supports these models.
Our observations suggest that these VLM cores may undergo some
active episodes of outflow and associated accretion with duration of a few $10^{3}$~yr.
The accretion rates, however, are very low, in the range 10$^{-11}$--10$^{-9}$~$M_{\odot}$yr$^{-1}$ or lower,
and they are comparable or smaller than the mass-loss rate.
This may therefore prevent the VLM cores to accrete enough material to become a star.

\section{SUMMARY}
Here, we report our latest search for molecular outflows
from VLM objects. So far, we have detected molecular outflows
from two BDs (one in $\rho$ Ophiuchi and one in Taurus) and one VLM star in Taurus.   
Our results suggest that: (1) the bipolar molecular outflow process in VLM objects 
is a scaled-down version of that in low-mass stars; 
(2) the outflow mass-loss and the associated mass
accretion processes in VLM objects are possibly episodic
with duration of a few thousand years;
(3) the outflow mass-loss rate and the mass accretion rate 
during active episodes are very low and they
do not significantly change for different stages of the formation process of VLM objects;
(4) A very low mass accretion rate, possibly together with  
a high ratio of outflow mass-loss rate to mass accretion rate, may prevent
a VLM core to accrete enough gas to become a star;
and thus the core will end up a BD.

\acknowledgments
This research is funded by Vietnam National Foundation for 
Science and Technology Development (NAFOSTED) under grant number 103.08-2013.21.
D.L. acknowledges the support from National Basic Research Program of China 
(973 program) No. 2012CB821800 and NSFC No. 11373038.
We thank the referee for valuable comments.
Support for CARMA construction was derived from the Gordon and Betty Moore Foundation, 
the Kenneth T. and Eileen L. Norris Foundation, the James S. McDonnell Foundation, 
the Associates of the California Institute of Technology, the University of Chicago, 
the states of California, Illinois, and Maryland, and the National Science Foundation. 
Ongoing CARMA development and operations are supported by the National Science Foundation 
under a cooperative agreement, and by the CARMA partner universities.

\clearpage

\clearpage

\clearpage

\begin{figure}
\vskip 1in
\hskip -0.25in
\centerline{\includegraphics[width=4in,angle=-90]{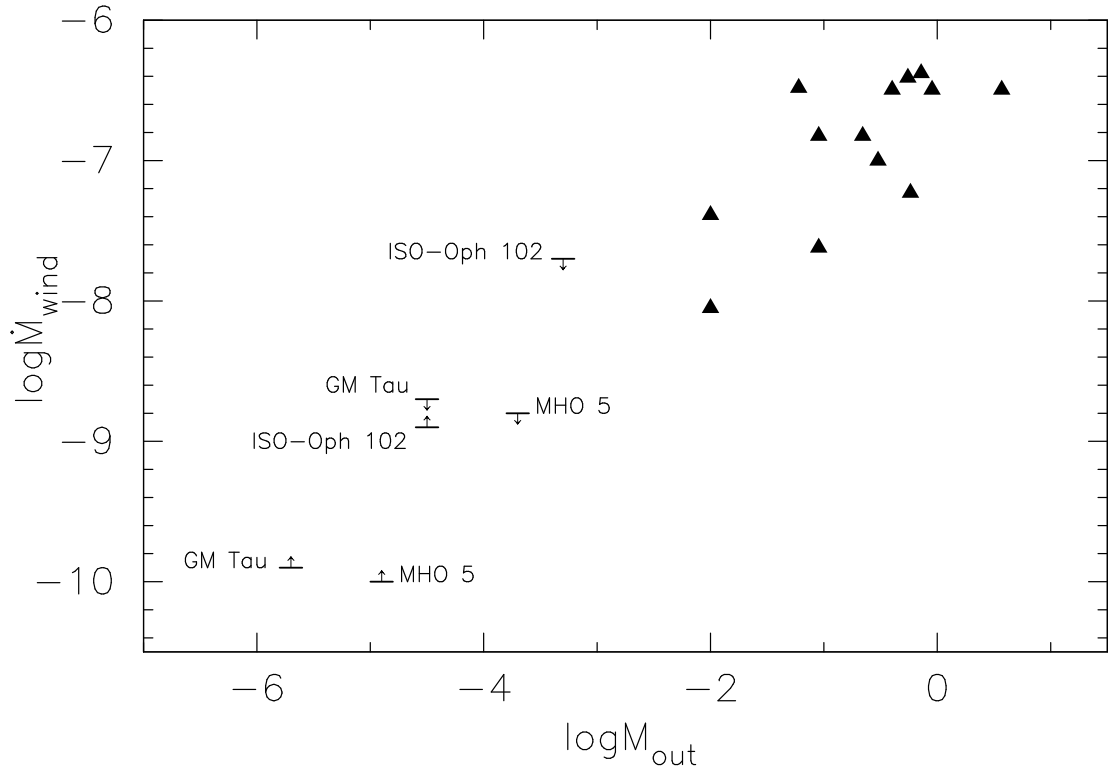}}
\caption{\normalsize Comparison between the molecular outflow mass and the wind mass-loss rate
of VLM objects and low-mass stars. The arrows up and down indicate the lower and upper
limits, respectively. Solid triangles represent class II low-mass stars with masses 
of 0.5--5.0~$M_{\odot}$
\citep{l88a}. 
\label{crys}}
\end{figure}

\clearpage

\begin{deluxetable}{llccccc}
\tabletypesize{\footnotesize}
\tablewidth{0pt}
\tablecaption{Observing logs for the three young BDs in Taurus
  \label{log}}
\tablehead { 
\colhead {Target} & \colhead {Array} & \colhead {Configuration} & \colhead {Beam size} & \colhead {continuum rms} & \colhead {line rms} & \colhead {velocity resolution}\\
\colhead {}       & \colhead {} & \colhead {} & \colhead {($\arcsec \times \arcsec$)}  & \colhead {(mJy/beam)}    & \colhead {(Jy/beam)}& \colhead {(km~s$^{-1}$)} \\
}
\startdata
 2M 0438      &  CARMA &  D       & 2.19$\times$1.68 &  1.4  &  0.20  & 0.11 \\
 GM Tau       &  SMA   &  Compact & 3.05$\times$2.82 &  1.0  &  0.15  & 0.27 \\
 2M 0414      &  CARMA &  D       & 2.09$\times$1.69 &  1.1  &  0.13  & 0.11 \\
\enddata
\tablecomments{For 2M~0438 and 2M~0414, line rms and velocity resolution are computed for bandwidth 31~MHz.}
\end{deluxetable}

\clearpage

\begin{deluxetable}{llclcccccccl}
\tabletypesize{\footnotesize}
\tablewidth{0pt}
   \tablecaption{Molecular outflow properties of class II VLM objects 
in $\rho$ Ophiuchi and Taurus \label{mloss}}
\tablehead {    
\colhead {Target} & \colhead {Array} & \colhead {Mass} & \colhead {Region} & \colhead {size (length)} & 
\colhead {$v_{\rm max}$} & \colhead {log$M_{\rm out}$\tablenotemark{a}}  & \colhead {log$\dot{M}_{\rm mol}$\tablenotemark{a}} &
\colhead {log$M_{\rm out}$\tablenotemark{b}}  & \colhead {log$\dot{M}_{\rm mol}$\tablenotemark{b}} & \colhead {$t_{\rm dyn}$} & 
\colhead {Reference}  \\
\colhead {}       & \colhead {}      & \colhead {($M_{\rm J}$)} & \colhead {} & \colhead {(AU)} & \colhead {(km~s$^{-1}$)} & 
\colhead {($M_{\odot}$)} & \colhead {($M_{\odot}$~yr$^{-1}$)} & \colhead {($M_{\odot}$)} & \colhead {($M_{\odot}$~yr$^{-1}$)} &
\colhead {(yr)} & \colhead {}      \\  
}
\startdata 
ISO-Oph 32   & SMA   & 40   & $\rho$ Oph & -   & -   &    -    &    -    &   -    &    -   &  -   & 1, 2  \\ 
ISO-Oph 102  & SMA   & 60   & $\rho$ Oph & 1000 & 4.7 & $-$4.5  & $-$7.5  & $-$3.3 & $-$6.3 &  1100 & 1, 3  \\
2M 0441+2534 & CARMA & 35   & Taurus     & -   & -   &     -   &    -    &   -    &    -   &  -   & 4, 2  \\
2M 0439+2544 & CARMA & 50   & Taurus     & -   & -   &     -   &    -    &   -    &    -   &  -   & 4, 2  \\
2M 0438+2611 & CARMA & 70   & Taurus     & -   & -   &     -   &    -    &   -    &    -   &  -   & 4, 5  \\
GM Tau       & SMA   & 73   & Taurus     &  ~700 & 4.6 & $-$5.7  & $-$8.6  & $-$4.5 & $-$7.4 & ~700 & 6, 5  \\
2M 0414+2811 & CARMA & 75   & Taurus     & -   & -   &     -   &    -    &   -    &    -   &  -   & 4, 5  \\
MHO 5        & SMA   & 90   & Taurus     &  ~600 & 2.1 & $-$4.9  & $-$8.3  & $-$3.7 & $-$7.1 &  2700 & 7, 2  \\
\enddata

\tablecomments{$^{\rm a}$Lower limits of outflow mass and mass-loss rate without mass corrections.}

\tablenotetext{b}{Upper limits of outflow mass and mass-loss rate with mass corrections: 
a factor of three for SMA missing flux \citep{bourke} and five for optical depth \citep{l88a}.}

\tablerefs{References for mass estimate, outflow mass and mass-loss rate: 
(1) \citet{natta04}; (2) \citet{pb11}; (3) \citet{pb08}; (4) \citet{muz05}; (5) this paper;
(6) \citet{white03}; (7) \citet{muz03}.}

\end{deluxetable}

\clearpage

\begin{deluxetable}{lccccl}
\tablewidth{0pt}
\tablecaption{Wind mass-loss rate of young VLM objects in $\rho$ Ophiuchi and Taurus
  \label{comp}}
\tablehead { 
\colhead {Target} & \colhead {log$\dot{M}_{\rm acc}$} & \colhead {log$\dot{M}_{\rm wind}$\tablenotemark{a}} & 
\colhead {log$\dot{M}_{\rm wind}$\tablenotemark{b}} &   \colhead {$\dot{M}_{\rm wind}$/$\dot{M}_{\rm acc}$\tablenotemark{c}}  & \colhead {Reference}  \\
 \colhead {}      &  \colhead {($M_{\odot}$~yr$^{-1}$)} & \colhead {($M_{\odot}$~yr$^{-1}$)} &  \colhead {($M_{\odot}$~yr$^{-1}$)} & \colhead {} & \colhead {} \\
}
\startdata
ISO-Oph 102  &  $-$9.0    &  $-$8.9  & $-$7.7 & 1.3-20.0     &     1  \\
GM Tau       &  $-$8.6    &  $-$9.9 & $-$8.7 & 0.05-0.8     &     2  \\
MHO 5        &  $-$10.8   &  $-$10.0 & $-$8.8 & 6.3-100.0    &     3  \\
\enddata
\tablecomments{$^{\rm a}$Lower limit of wind mass-loss rate.}
\tablenotetext{b}{Upper limit of wind mass-loss rate.}
\tablenotetext{c}{The range of the ratio of wind mass-loss rate to the accretion rate.}
\tablerefs{References for accretion rate: (1)  \citet{natta04}; (2) \citet{white03}; (3) \citet{muz03}.}
\end{deluxetable}

\end{document}